\documentclass[acmtaco]{acmsmall}

\usepackage{mathtools}
\usepackage{tikz}
\usepackage{cite}
\usepackage{xcolor}
\usepackage[textsize=tiny, textwidth=1.75cm]{todonotes}
\usetikzlibrary{calc}
\usetikzlibrary{shapes,snakes}
  \usetikzlibrary{positioning,arrows}
\tikzstyle{myarrows2}=[line width=1mm,draw=black,->] 
\tikzstyle{myarrows}=[line width=1mm,draw=blue,-triangle 45, postaction={draw, line width=1.5mm, shorten >=0.3mm, -}]
\tikzstyle{myarrows1}=[line width=1mm,draw=red,  dashed, postaction={draw, dashed, line width=1.1mm}]
\tikzstyle{myarrows3}=[line width=0.5mm, dashed, draw=black,-] 
\newcommand{\commentout}[1]{}

\usepackage{graphicx,caption}
\usepackage[utf8]{inputenc}
\usepackage[T1]{fontenc}

\usepackage{url}

\begin{document}

\title{Scheduling and Tiling Reductions on Realistic machines}

\author{Nirmal Prajapati\\ 
October 25, 2016}

 
\date{October 25, 2016}


\begin{abstract}
Computations, where the number of results is much smaller than the input data and are produced through some sort of accumulation, are called Reductions.  Reductions appear in many scientific applications.  Usually, reductions admit an associative and commutative binary operator over accumulation.  Reductions are therefore highly parallel.  Given unbounded fan-in, one can execute a reduction in constant/linear time provided that the data is available.  However, due to the fact that real machines have bounded fan-in, accumulations cannot be performed in one time step and have to be broken into parts.  Thus, a (partial) serialization of reductions becomes necessary.  This makes scheduling reductions a difficult and interesting problem. 

There have been a number of research works in the context of scheduling reductions.  We focus on the scheduling techniques presented in ~\cite{Gupta}, identify a potential issue in their scheduling algorithm and provide a solution.  In addition, we demonstrate how these scheduling techniques can be extended to ``tile" reductions and briefly survey other studies that address the problem of scheduling reductions.
\end{abstract}

\maketitle

\section{Introduction}

Reductions are those computations in which an associative and commutative operator accumulate a set of points into a single value.  Following example illustrates such a computation. 

\begin{verbatim}
   for i = 1 to N 
            A[i] += B[i-1];
\end{verbatim}

The reduction operator is associative and commutative, which implies that accumulations need not admit any order. Therefore, we should be able to exploit parallelism.  With an unbounded number of processors and unbounded fan-in operator, accumulations can be done in a single time step.  However, for a real machine, both the number of processors and the fan-in are bounded. This necessitates ordering of accumulations.

Many scientific and engineering applications spend most of their execution time in nested loops.  The task of optimizing such nested loops involve dataflow analysis of the program~\cite{feautrier91}.  
The dataflow analysis of the above program reflects loop-carried data dependences that prevent parallelism.  Therefore, scheduling reductions is a difficult problem.

A multidimensional affine function that imposes a particular order of execution is called a Schedule.   A schedule must satisfy the precedence constraints imposed by the dependences.  Polyhedral model renders a powerful abstraction that enables precise reasoning for the legality of transformations.  Iterations of the nested loops can be viewed as integer points in a polyhedron.  The computations impose dependence constraints which are represented as affine functions of the indices.  Reductions can also be described as Systems of Affine Recurrence Equations (SAREs) over polyhedral domains.  Reduction dependences are implicit in the SAREs.  Such a representation allows us to focus on non-reduction dependences that must be satisfied by the schedule.

~\cite{Redon} present a scheduling technique that optimally schedules reductions over CRCW PRAM model.  They assumed that accumulations can happen in one time step.  This approach was extended by ~\cite{Gupta} to work on realistic machines.  They invented a scheduling technique for machines with binary operators and exclusive writes.  They claim that on such machines, their scheduling method gives efficient solutions with a constant factor slowdown compared to best possible schedules on a CRCW PRAM model.  We discover a flaw in their technique which violates their claim of ``exclusive writes".  Using a counter example, we prove that their scheduling technique requires a machine with concurrent writes.  We solve this problem by introducing additional causality constraints and show that the modified scheduling constraints guarantee a schedule with exclusive writes.

Furthermore, we extend the scheduling approach to tile reductions.  Tiling~\cite{Wol87,Irigoin1988} is a classic iteration space partitioning technique which combines a set of points into tiles, where each tile can be executed atomically.  Tiling comes in handy for exploiting data locality~\cite{WL-pldi91}, minimizing communication ~\cite{Andonov2001,xue-jpdc97} and maximizing parallelism.  Reductions are usually serialized before tiling.  Most of the tiling techniques such as~\cite{uday-pldi08,Polly} take as input serialized reduction programs.  Loop transformation techniques are used to find tiling hyperplanes.  In the majority of the cases, serializing imposes uniform dependences which are tileable.  Serializing reductions, however, negates the fact that accumulations can be carried out in any order and imposes unnecessary intra-tile and inter-tile dependences.  The tiling techniques developed in this paper maximize parallelism as well as improve data locality in a work efficient manner.  

We also discuss other limitations of Gupta's scheduling algorithm and suggest possible improvements.  We highlight the unexplored areas and address other related work in the context of scheduling reductions.  

The rest of the paper is organized as follows.  Section 2 describes necessary background.  In section 3, we describe~\cite{Gupta} scheduling technique using examples.  Section 4 exposes the flaw in their technique using a counter example which shows that ``exclusive writes" condition is violated and proposes a solution that is guaranteed to give schedules for machines with ``exclusive writes".  Section 5 extends this scheduling technique to tile reductions.  Section 6 suggests other possible improvements, section 7 discusses related work and section 8 concludes the paper.


\section{Background}\label{background}

A reduction of variable $X$ can be represented as 
\begin{align}
X = reduce(\oplus, f, R)
 \label{eq:X}
\end{align}
where $\oplus$ is an associative and commutative accumulation operator, the body of the reduction is some variable $R$, and $f$ is a projection function that maps a subset of the points in the domain of $R$, represented as $Dom(R)$, to $z_{X} \in Dom(X)$.  For every $z_{X} \in Dom(X)$, the \emph{Parametrized Reduction Domain} of $z_{X}$, referred as $P(z_{X})$, is the set of points in $R$ that are mapped to $z_{X}$ by the function $f = z \rightarrow 
           		A_{P}
           		\begin{pmatrix}
           		 z \\           
          		 p \\
           		\end{pmatrix}
           		+ c_{P} $
where $A_{P}$ is a constant matrix, $c_{P}$ is a constant vector and $p$ is a vector of the size parameters.

\subsection{Schedule}
Schedule of a variable $X$ is a vector that represents the time instant at which $z_{X} \in Dom(X)$ is computed and is given by $\lambda_{X}(z_{X})$, a multidimensional affine function.
For any two variables $X$ and $Y$, if $X(z)$ depends on $Y(z')$ then $Y(z')$ must be computed before $X(z)$. 
The dependence imposes the following causality constraints on the schedule

\begin{align}
 \lambda_{X}(z) \succeq \lambda_{Y}(z') + T_{eqX}(z, z')
 \label{eq:schedule}
 \end{align}
 
where $T_{eqX}(z, z')$ is the time to compute the RHS of the equation $X(z)$ after $Y(z')$ becomes available.

~\cite{Redon} scheduling algorithm assumes a CRCW PRAM with unbounded number of processors and unbounded fan-in operators. The reductions can, therefore, be performed in a single time step.  The causality constraints for equation (\ref{eq:X}) on such a machine are given by

\begin{center}
$z_{X} \in Dom(X),  z_{R} \in P(z_{X}), $
\end{center}
\begin{align}
 \lambda_{X}(z_{X}) \succeq \lambda_{R}(z_{R}) + 1
 \label{eq:scheduleX}
\end{align}

However, this technique is not applicable for scheduling reductions on real machines.  ~\cite{Gupta} developed a technique that schedules reductions on realistic machines with bounded fan-in (binary) operators and exclusive writes.  They claim that their scheduling algorithm generates efficient schedules with a constant fold slow down compared to optimal schedules obtained on a CRCW PRAM model.  

The following section explains in detail the scheduling algorithm as presented in ~\cite{Gupta}.

\section{Gupta's scheduling algorithm}\label{gupta}
Let $\lambda_{R}(z_{R})$ be the schedule of $z_{R} \in Dom(R)$.
$\lambda_{R}(z_{R}) = t$ are the \emph{equitemporal hyperplanes}, or ``slices", defined as the set of points in $z_{R} \in P(z_{X})$ that become available for accumulation at time $t$. A temporary variable $TempX(z_{X}, t)$ is defined as follows  

\begin{align}
    Temp_{X} &= reduce
        \begin{pmatrix}
        \oplus,
        	\begin{pmatrix}
        	z \rightarrow 
        		\begin{pmatrix}
           		A_{P} \\           
          		 \Lambda_{R} \\
           		\end{pmatrix}
           		\begin{pmatrix}
           		 z \\           
          		 p \\
           		\end{pmatrix}
           		+
           		\begin{pmatrix}
           		c_{P} \\           
          		 \alpha_{R} \\
           		\end{pmatrix}
           \end{pmatrix}
           , R
    \end{pmatrix}
    \label{eq:TempX}
\end{align}

where  $ \lambda_{R}(z_{R}) =    
          		 \Lambda_{R}
           		\begin{pmatrix}
           		 z_{R} \\           
          		 p \\
           		\end{pmatrix}
           		+ \alpha_{R} $
 is the schedule for $R$.

$TempX(z_{X}, t)$ are the partial accumulations of equitemporal hyperplanes in $P(z_{X})$. These intermediate results are accumulated to get final answer $z_{X} \in Dom(X)$.  Equation (\ref{eq:X}) is modified as

\begin{align}
X = reduce(\oplus, (z_{X}, t \rightarrow z_{X}), TempX)
 \label{eq:guptaX}
\end{align}

Consider, $X$ as the following reduction:
\begin{equation}
X = \sum_{i=0, j=0}^{i=N, j=i} R(i, j)
\label{eq:guptaeg}
\end{equation}
\begin{center} $Dom(R)=\{i, j | 0\le j\le i \le N\}$ \end{center}

\begin{figure}
\centering
\includegraphics[scale=0.5]{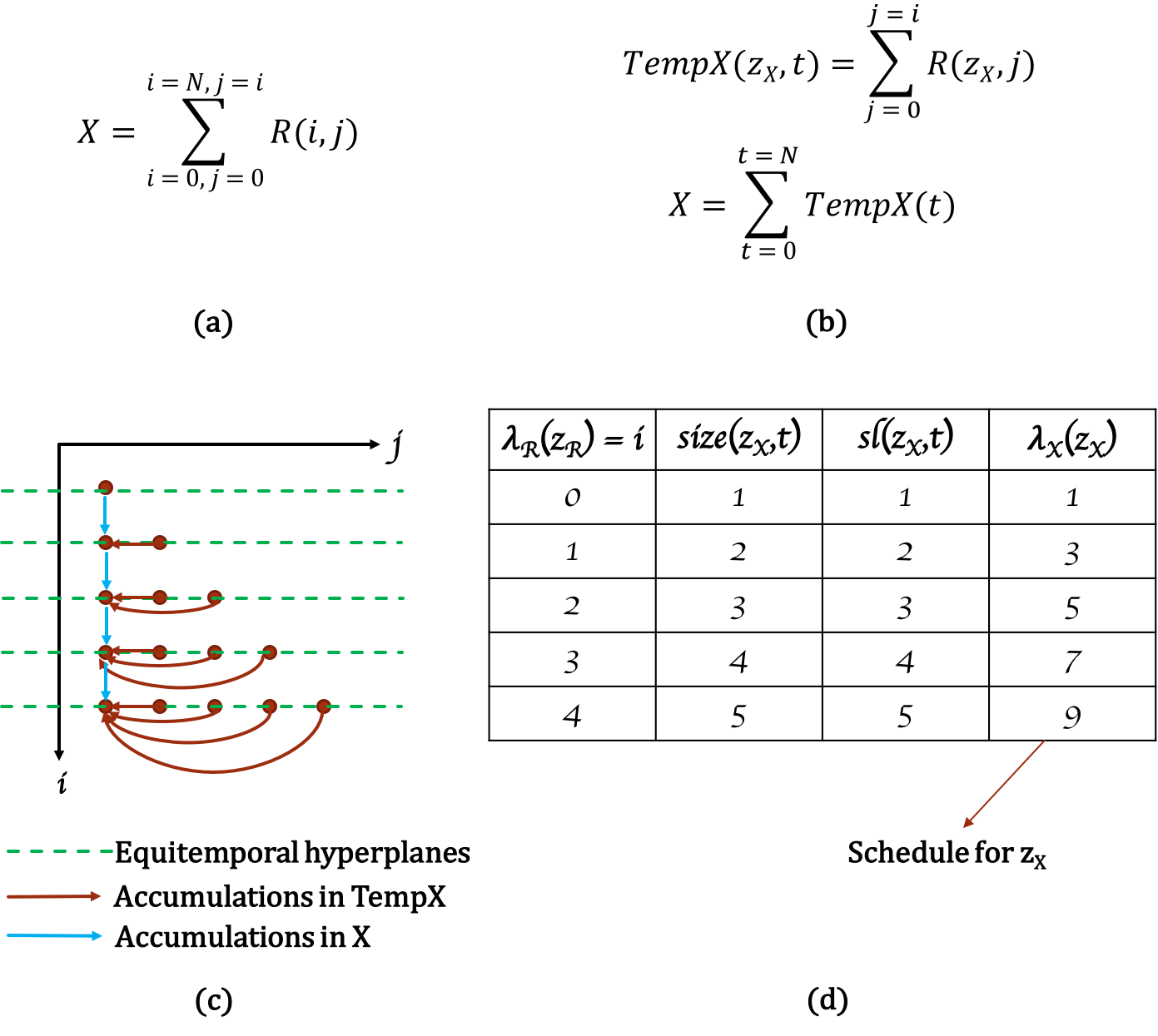}
\caption{Gupta's scheduling technique for the equation in (a).}
\label{fig:technique}
\end{figure}

Assume, $\lambda_{R}(z_{R}) = i$ are the equitemporal hyperplanes. With this information, we can deduce that there are $N$ equitemporal hyperplanes. i.e. $Dom(TempX)$ has $N$ elements where the $t$-th element is a partial accumulation of the $t$-th equitemporal hyperplane in $P(z_{X})$.  Figure~\ref{fig:technique}(b) shows the new set of equations obtained after decomposing the reduction as shown in equations (\ref{eq:TempX}) and (\ref{eq:guptaX}).

Correspondingly, the causality constraints in equation (\ref{eq:schedule}) are updated to accommodate $TempX$.
\begin{center}
$f(z_{X}) \leq t \leq l(z_{X}),$
\end{center}
\begin{align}
\lambda_{TempX}(z_{X},t) \succeq t + T_{eqTempX}(z_{X},t) \\
\lambda_{X}(z_{X}) \succeq \lambda_{TempX}(z_{X},t) + T_{eqX}(z_{X},t) 
 \label{eq:guptaScheduleX}
\end{align}

where $f(z_{X})$ and $l(z_{X})$ are the first and last time steps at which values in $\mathcal{P}(z_{X})$ are available.  $T_{eqTempX}(z_{X},t)$ is the time to compute the reduction of all the values in the equitemporal hyperplane $(z_{X},t)$.  For a machine with binary operators, $size(z_{X},t)$ gives the time to accumulate the equitemporal hyperplane at $(z_{X},t)$. The number of time steps required for linear accumulation of all values in any box B is defined as its $size$, which is also equal to the 1-norm of its principal diagonal.

For the example in Figure~\ref{fig:technique}(a), the $size(z_{X}, t)$ of a hyperplane is given by $t = i$.  Figure~\ref{fig:technique}(c) shows the iteration space and the table in Figure~\ref{fig:technique}(d) give the sizes of the equitemporal hyperplanes at $(z_{X},t)$.

The scheduling constraints in equations (7) and (\ref{eq:guptaScheduleX}) get reduced to 
\begin{align}
size(z_{X},t) \preceq \lambda_{X}(z_{X}) - t 
 \label{eq:guptasizeX}
\end{align}

The \emph{slack} of an equitemporal hyperplane is defined as $sl(z_{X},t) = \lambda_{X}(z_{X}) - t$.   The scheduling constraints in (\ref{eq:guptasizeX}) are further modified to

\begin{align}
sl(z_{X},t) \succeq size(z_{X},t) 
 \label{eq:guptaslackX}
\end{align}

Note that by definition $size(z_{X},t)$ is always one-dimensional scalar. If slack is truly multidimensional then the constraint in (\ref{eq:guptaslackX}) is trivially satisfied. However, if the slack is not multidimensional, then the value of the innermost dimension of slack should be greater than the $size(z_{X},t)$.

These causality constraints can be formulated as an integer linear program and solved using a PIP~\cite{PIP} solver.  Using the above constraints to formulate causality for the example in Figure~\ref{fig:technique}, we obtain schedule $\lambda_{X}(z_{X})$ as shown in Figure~\ref{fig:technique}(d).

\begin{figure}
\centering
\includegraphics[scale=0.55]{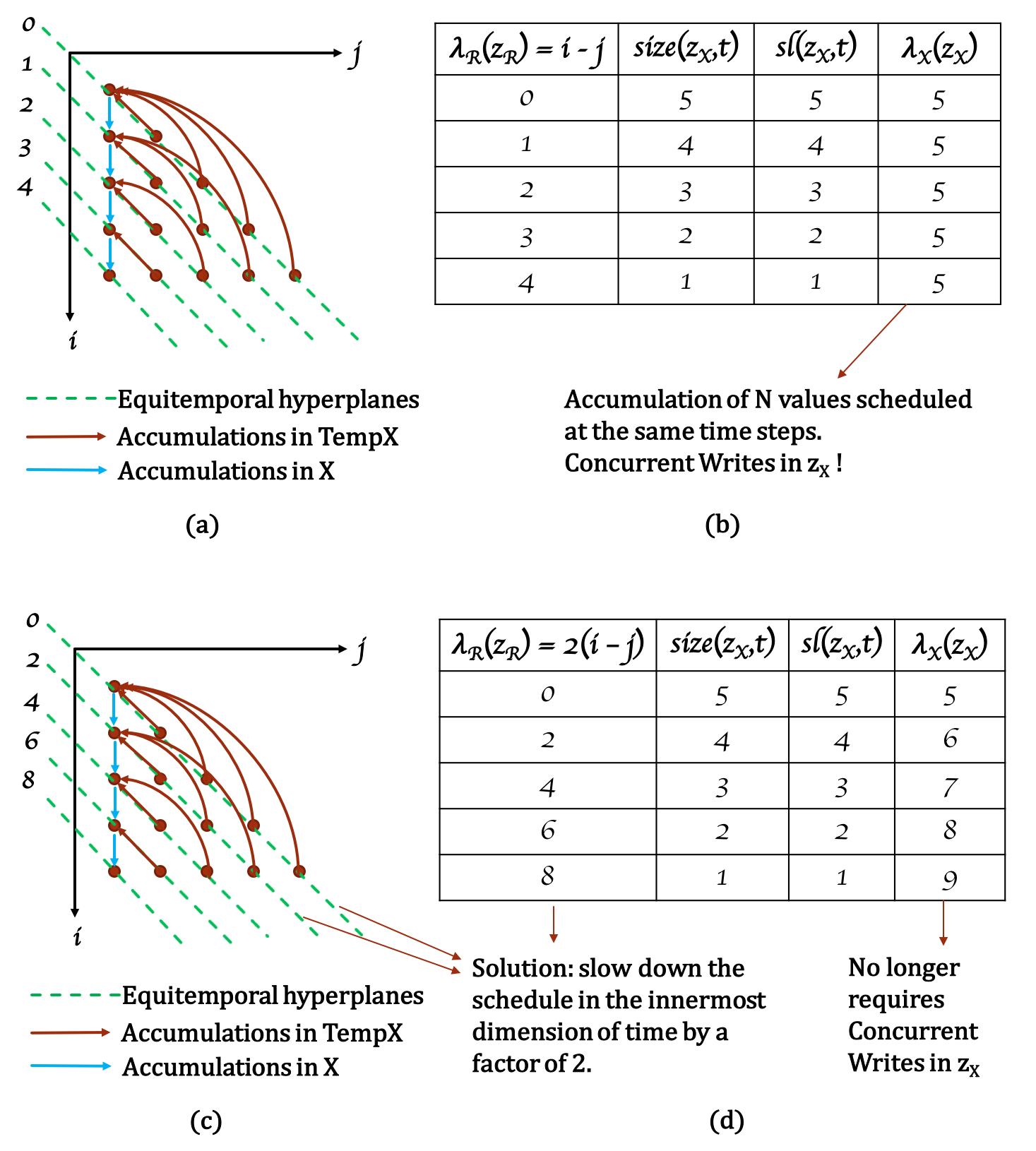}
\caption{Gupta' scheduling technique for equation (\ref{eq:guptaeg}), given $\lambda_{R}(z_{R}) = i-j$.}
\label{fig:technique2}
\end{figure}

For the same example, assume that $\lambda_{R}(z_{R}) = i-j$ are the equitemporal hyperplanes.  Figure~\ref{fig:technique2}(a) shows the hyperplanes at which values in $P(z_{X})$ become available for accumulation.  For this example, $P(z_{X}) = Dom(R)$.  Again, there are $N$ equitemporal hyperplanes. i.e. $Dom(TempX)$ has $N$ elements where the $t-th$ element is a partial accumulation of the $t-th$ equitemporal hyperplane in $z_{X}$.

The $size(z_{X}, t)$ of each hyperplane is shown in Figure~\ref{fig:technique2}(b).  We obtain the schedule for $z_{X}$ such that it requires a machine with concurrent writes.  As shown in Figure~\ref{fig:technique2}(b), the accumulations are scheduled at the same time step.  In such cases, ~\cite{Gupta} suggests that we slow down the schedule by a factor of 2 in the innermost dimension of time and obtain enough time for the accumulations in equation (\ref{eq:guptaX}) on a machine with binary operators.  Figures~\ref{fig:technique}(c) and (d) show how this modification easily solves the problem.

In the next section, we provide a counter example to show that this modification does not solve the problem of concurrent writes in all cases.

\section{Counter Example}\label{sec:solution}

Consider, variable $X$ as the following reduction equation:
\begin{equation}
X(i) = \sum_{j=0}^{j=i} R(i, j)
\label{eq:guptaeg2}
\end{equation}
\begin{center} $Dom(R)=\{i, j | 0\le j\le i \le N\}$ \end{center}

\begin{figure}
\centering
\includegraphics[scale=0.55]{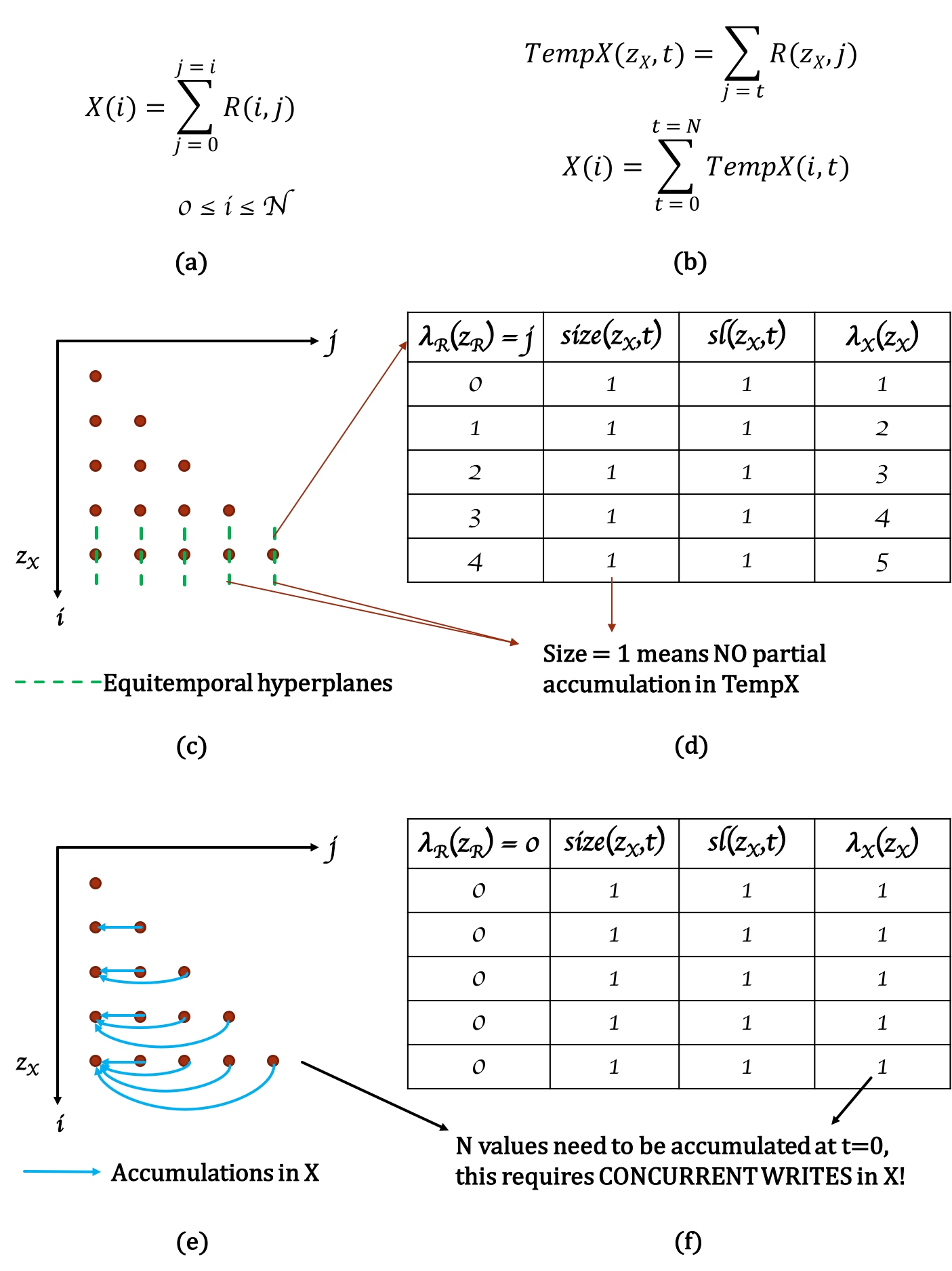}
\caption{Gupta's scheduling technique violates \emph{Exclusive Write} condition for equation (\ref{eq:guptaeg2}).}
\label{fig:technique2-flaw}
\end{figure}

Here, $X$ is a one-dimensional variable. $Dom(R)$ is again two-dimensional. Assume, $\lambda_{R}(z_{R}) = j$ are the equitemporal hyperplanes. With this information, we can see that there are $j$ equitemporal hyperplanes. i.e. the slice at $(z_{X}, t)$ has a single element. The $t-th$ element is a partial accumulation of the $t-th$ equitemporal hyperplane in $z_{X}$ which is a single point, therefore, there are no partial accumulations.

The size of all the equitemporal hyperplanes is equal to $1$.  Figure~\ref{fig:technique2-flaw}(c) shows the iteration space of equation (\ref{eq:guptaeg2}) and  Figure~\ref{fig:technique2-flaw}(d) shows the desird schedule.

Using these constraints, we will now solve for $t$.  Assume, we get $\lambda_{R}(z_{R}) = t = 0$. The schedule $\lambda_{X}(z_{X})$ is calculated using the constraint in equation (\ref{eq:guptaslackX}).  This constraint is trivially satisfied with a size of 1 for all equitemporal hyperplanes.  Figures~\ref{fig:technique2-flaw}(e) and (f) show the issue.  Here the schedule is zero-dimensional and we can no longer slow it down by a constant factor to accumulate the intermediate results.  This violates the \emph{``exclusive write"} condition and requires a machine with \emph{Concurrent Writes}!.

In the previous example, note that the number of values to be accumulated into $z_{X}$ is more than then number of time steps between $f(z_{X})$ and $l(z_{X})$.  This prevents linear accumulation. 
We define $size'(z_{X})$ as the total number of time steps required for linear accumulation of all values in the reduction body $TempX(z_{X}, t)$ of $X$ in equation (\ref{eq:guptaX}).  Let the total number of time steps between $f(z_{X})$ and $l(z_{X})$ be $T_{z_{X}}$.

If the condition 
\begin{equation}
\forall z_{X} \in Dom(X), T_{z_{X}} + 1 > size'(z_{X})
\label{eq:total}
\end{equation}

is satisfied, then concurrent writes in $z_{X}$ can be avoided.  Suppose, $f(z_{X})=1$ and $l(z_{X})=N$.  Therefore, $T_{z_{X}}=N$.  If $size'(z_{X}) < N$, then the condition is trivially satisfied.  However, this constraint cannot be satisfied for the example (\ref{eq:guptaeg2}) where $T_{z_{X}} = 1$ and $size'(z_{X}) = N$.  

Let the total number of equitemporal hyperplanes in $P(z_{X})$ be $E_{z_{X}}$.   Again, if  $E_{z_{X}} < size'(z_{X}) < T_{z_{X}}$, then exclusive writes cannot be guaranteed by equation (\ref{eq:total}).

Now let's see what happens if the condition 
\begin{equation}
\forall z_{X} \in Dom(X), E_{z_{X}} + 1 > size'(z_{X})
\label{eq:totalhyperplanes}
\end{equation}
is satisfied.  There would be enough time steps for the accumulation of the intermediate values into the final answer $z_{X}$.  However, if there is only one equitemporal hyperplane like our example (\ref{eq:guptaeg2}), then the constraint in (\ref{eq:totalhyperplanes}) will not be satisfied.  Neither the number of equitemporal hyperplanes $E_{z_{X}}$ nor the total number of time steps $T_{z_{X}}$ can be guaranteed to be more than $size'(z_{X})$.

Therefore, we suggest the following scheduling constraints 
\begin{align}
\lambda_{X}(z_{X}) \succeq size'(z_{X})
 \label{eq:goodscheduleX}
\end{align}
in addition to the constraints in (\ref{eq:guptaslackX}).  Linear accumulations are now guaranteed on a machine with bounded fan-in.

The above scheduling technique can be further optimized to get better schedules.  In the next section, we will see how this scheduling technique can be extended to tile reductions in order to maximize parallelism and improve data locality.

\section{Tiling Reductions}\label{tilingre}
Tiling or blocking computations is a strategy of dividing the iteration space into tiles where each tile is a set of points~\cite{Wol87,Irigoin1988}.  A tiling is considered to be legal if there are no dependence based cycles between tiles and if all tiles can be executed atomically.

Let $\theta_{X}(\overrightarrow{i})$ define a set of tiling hyperplanes that tile the iteration space of a variable $X$.
For any two variables $X$ and $Y$, if $X$ depends on $Y$ then the following tiling legality constraint must be satisfied for all the dependencies between $X$ and $Y$

  \begin{align}
  \theta_{X}(\overrightarrow{i}) - \theta_{Y}(\overrightarrow{i}) \geq 0
  \label{eq:tilinglegality}
  \end{align}

The above condition ensures the legality of tiling as shown in~\cite{Bondhugula2008}.

With the knowledge that accumulations can be carried out in any order, we can eliminate the reduction dependences from the dependence set.  However, (\ref{eq:tilinglegality}) must hold for all other dependencies.

\begin{figure}
\centering
\includegraphics[scale=0.55]{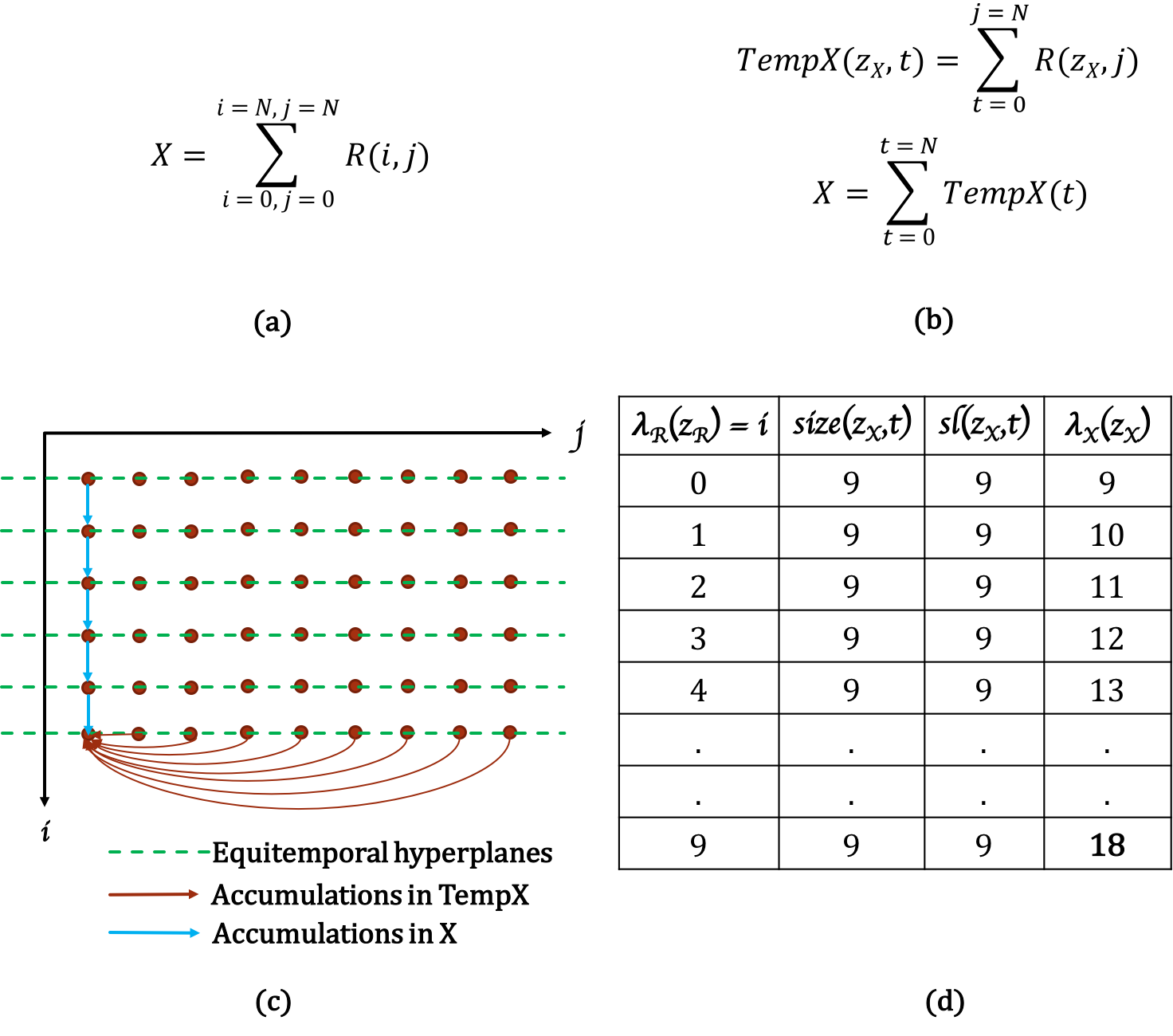}
\caption{~\cite{Gupta} scheduling technique for equation in (a), given $\lambda_{R}(z_{R}) = i$.}
\label{fig:tile-eg}
\end{figure}

Consider the equation shown in Figure~\ref{fig:tile-eg}(a).  Figure~\ref{fig:tile-eg}(b) shows the decomposition of $X$ as per equation (\ref{eq:TempX}).  Figures~\ref{fig:tile-eg}(c) and (d) show the iteration space and schedule obtained using the formulation discussed in Section~\ref{gupta}.

We provide an incremental approach to finding tiling hyperplanes for reductions. We first show how equitemporal hyperplanes can be tiled, succeeded by tiling $P(z_{X})$ and finally suggest possible tilings for the reduction body $R$.

\subsection{Tiling Equitemporal Hyperplanes}\label{tilingequi}

Tiling an equitemporal hyperplane $(z_{X},t)$ using any tiling hyperplane is a legal tiling.  This is due to the fact that all values in an equitemporal hyperplane are available for accumulation at the same time and that all of them contribute to a single value in $TempX$.  Therefore, the tiling legality condition (\ref{eq:tilinglegality}) holds for any tiling hyperplane.  We are left with many possible choices.  We choose orthogonal tiling hyperplanes with tiles of size $s$ in every dimension.

We introduce a new variable $Tile_{TempX}$ such that $(z_{X}, t, b) \in Dom(Tile_{TempX})$ maps to the $b$th tile in $TempX$, where $(z_{X},t) \in Dom(TempX)$.

\begin{align}
    Tile_{TempX} &= reduce
        \begin{pmatrix}
        \oplus,
        	\begin{pmatrix}
        	z \rightarrow 
        		\begin{pmatrix}
           		A_{P} \\           
          		 \Lambda_{R} \\
          		 \gamma
           		\end{pmatrix}
           		\begin{pmatrix}
           		 z \\           
          		 p \\
          		 s \\
           		\end{pmatrix}
           		+
           		\begin{pmatrix}
           		c_{P} \\           
          		 \alpha_{R} \\
           		\end{pmatrix}
           \end{pmatrix}
           , R
    \end{pmatrix}
        \label{eq:TileTempX}
  \end{align}
  
where $\gamma$ is a function that divides every dimension of the equitemporal hyperplane at $(z_{X},t)$ into tiles of size $s$ such that $1 \leq b \leq \tau(z_{X},t)$, where $\tau(z_{X},t)$ is the total number of tiles in the \emph{slice} at $(z_{X},t)$.  With this definition of variable $Tile_{TempX}$, equation (\ref{eq:TempX}) is modified as
  
  \begin{align}
  TempX = reduce (\oplus, (z_{X},t,b \rightarrow z_{X}, t), Tile_{TempX})
  \label{eq:reduceTempX}
  \end{align}
  
$TempX(z_{X}, t)$ is the accumulation of $\tau(z_{X}, t)$ tiles. Equation (\ref{eq:guptaX}) remains unchanged.

\begin{figure}
\centering
\includegraphics[scale=0.5]{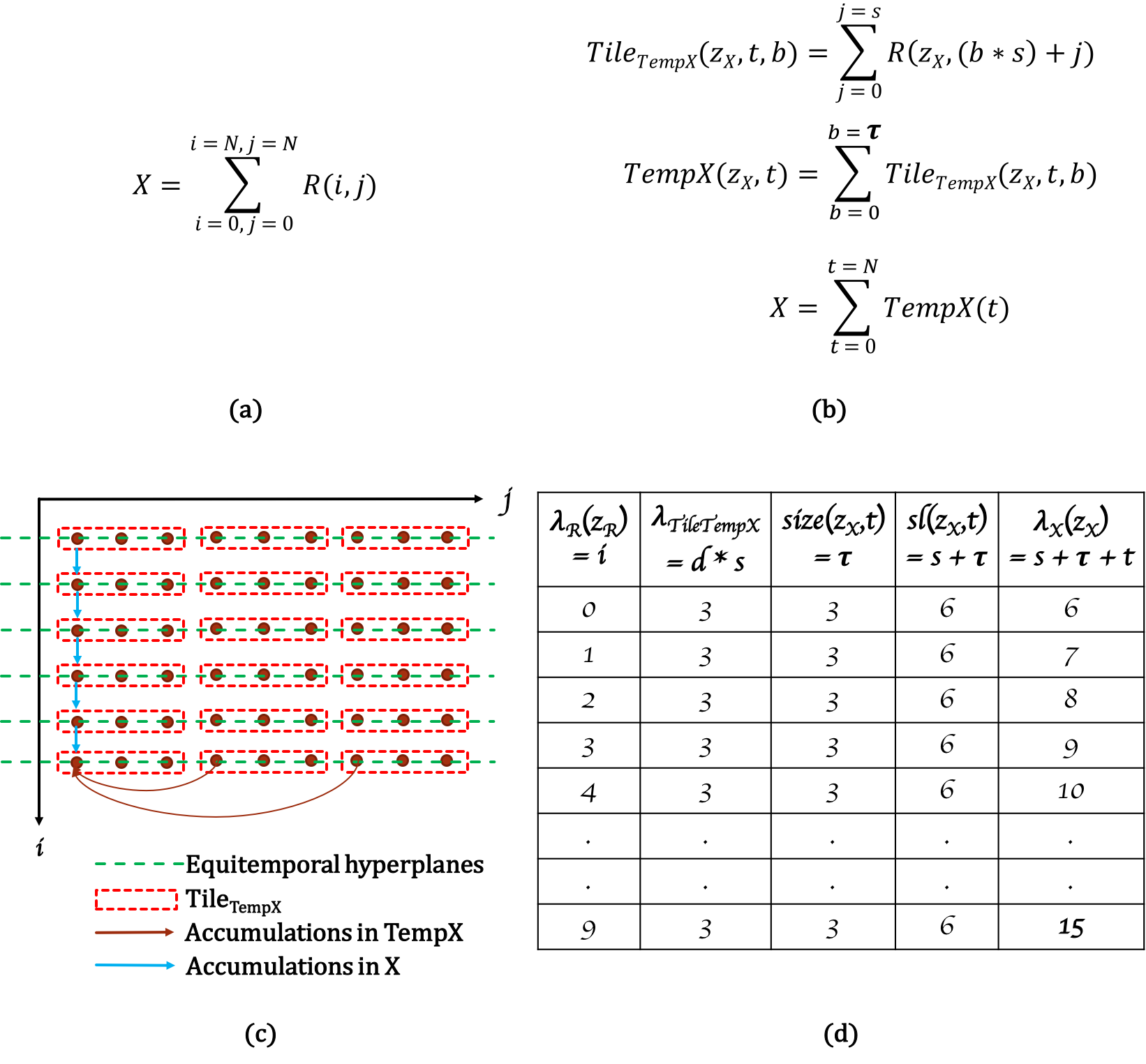}
\caption{Extending the scheduling techniques for tiling equitemporal hyperplanes for the equation in (a), given $\lambda_{R}(z_{R}) = i$.}
\label{fig:tile-example}
\end{figure}

Consider the equation shown in Figure~\ref{fig:tile-example}(a).  Figure~\ref{fig:tile-example}(b) shows the decomposition of $X$ as per equations (\ref{eq:TileTempX}) and (\ref{eq:reduceTempX}).  We now obtain causality constraints for equations (\ref{eq:TileTempX}) and (\ref{eq:reduceTempX}).  The precedence constraints on $Tile_{TempX}$ state that 
\begin{center}
$1 \leq b \leq \tau(z_{X},t), f(z_{X}) \preceq t \preceq l(z_{X}), zx \in Dom(X)$
\end{center}
\begin{align}
  \lambda_{Tile_{TempX}}(z_{X},t,b) \succeq t + T_{eqTile_{TempX}}(z_{X},t,b) \\
   \lambda_{TempX}(z_{X},t) \succeq \lambda_{Tile_{TempX}}(z_{X},t,b) + T_{eqTempX}(z_{X},t)
\end{align}

In an equitemporal hyperplane, all the tiles can be executed simultaneously.  $T_{eqTile_{TempX}}(z_{X},t,b)$ is given by the size of the tile.  We assume that tile size is $s$ in every dimension.  Let the number of dimension of the equitemporal hyperplane be $d(z_{X},t)$.  Hence, $size$ of a tile $(z_{X},t,b)$ will be given by $d(z_{X},t) \times s$.
We get
\begin{align}
  \lambda_{Tile_{TempX}}(z_{X},t,b) \succeq t + [ d(z_{X},t)\times s ]
    \label{eq:scheduleTile-e}
\end{align}
as the constraints for $\lambda_{Tile_{TempX}}(z_{X},t,b)$.

$T_{eqTempX}(z_{X},t)$ of $TempX$, defined as the $size(z_{X},t)$, is the time it takes to accumulate all the partial answers produced by each tile, which is also equal to $\tau(z_{X},t)$. Therefore, causality constraints on the schedule for equation (\ref{eq:reduceTempX}) can be formulated as:

\begin{align}
  \lambda_{TempX}(z_{X},t) \succeq t + \tau(z_{X},t) + [ d(z_{X},t)\times s ]
  \label{eq:scheduleTile-eTempX}
\end{align}

Similarly, we deduce the following as the causality constraints for equation (\ref{eq:guptaX}).
\begin{align}
  \lambda_{X}(z_{X}) \succeq t + \tau(z_{X},t) + [ d(z_{X},t)\times s ] \\
  \lambda_{X}(z_{X}) \succeq size'(z_{X})
  \label{eq:scheduleTile-eX}
\end{align}

  Figures~\ref{fig:tile-example}(c) and (d) show the iteration space and schedule obtained using the formulation discussed above. Using tile size $s = 3$ and $N = 9$, we get the number of tiles $\tau = 3$ and hence $\lambda_{X}(z_{X})$ can be scheduled as early as $15$.

Note, when $s$ does not equally divide every dimension of an equitemporal hyperplane, we get partial tiles whose size is smaller than full tiles. Therefore, above causality constraints are satisfied trivially for partial tiles.

With the above formulation, there are many possible choices for tile size $s$.  If we choose $s = 1$, then there will be only one point in each tile.  We provide a cost function that leads to good tile size and maximizes parallelism. 

\subsubsection{Towards finding good solutions}

To minimize the total execution time, reductions of an equitemporal hyperplane can be implemented using a binary-tree-like algorithm. However, this is not work-efficient.  To accumulate $N$ elements, binary-tree algorithms take $log N$ steps.  Brent's theorem suggests $N/log N$ parallel instances where each instance performs $log N$ work~\cite{brent99}.  Later, all $N/log N$ parallel instances contribute to the final accumulation of the partial results. The cost is now given by $(N/log N) * log N = N$, which is work efficient.  Our proposed cost function provides such work efficient solutions. 

We seek to minimize the following cost function
Let, 
\begin{center}
$\omega = [ d(z_{X},t)\times s ] - \tau(z_{X},t)$ 
\end{center}
\begin{align}
   minimize(\omega)
  \label{eq:omega}
\end{align}

The tile size $s$ used in Figures~\ref{fig:tile-example}(c) and (d) reflect the result of the cost function (\ref{eq:omega}).

Note that our tile size optimization function assumes equal tile size in every dimension. This restriction can be lifted to enable rectangular tiling.

This technique of tiling equitemporal hyperplanes applies to those cases where the equitemporal hyperplane is at least one dimensional.  If the hyperplanes are zero-dimensional then above formulation does not apply.  This motivates tiling across equitemporal hyperplanes.

\subsection{Tiling Parametrized Reduction Domain of \textbf{$z_{X}$}} \label{tilingpzx}
As shown above, tiling equitemporal hyperplanes is straightforward.  Let us consider tiling the Parameterized reduction domain of $z_{X}$.  If there are no dependencies between equitemporal hyperplanes, then orthogonal tiling hyperplane can be chosen and all tiles can be launched independently. The partial answers can be accumulated to get the final answer $z_{X}$.

However, if there exists dependence in $t$, then orthogonal tiling hyperplanes might not be legal.  Notice, the only dependences that affect tiling legality are between equitemporal hyperplanes (as per the definition of slices, all values become available for accumulation in an equitemporal hyperplane at the same time).  The problem of tiling $P(z_{X})$ is thus reduced to time tiling.  A well-known approach is to enforce forward communication only constraint as presented in ~\cite{Griebl2005}.  If we want to maximize parallelism and minimize synchronization, then we can use the time-partitioning technique of ~\cite{min-sync-lim-lam}.  If minimizing communication is also desired optimization, then the cost optimization techniques of ~\cite{uday-pldi08} can be used. If the dependences are uniform, then time tiling techniques for stencil computations such as~\cite{Tang2011,grosser-etal-GPUhextile-CGO2014,Diamond2016} can be used for maximizing parallelism together with the concurrent start.  

The dependences in $t$ impose additional constraints on tile sizes.  The tile size along each tiling hyperplane must be greater than the length of the longest dependence in the hyperplane.  Using these tiling hyperplanes and additional dependence based constraints, we can now optimize for tile size with our cost function. 

We can eliminate the variable $TempX$ and use only one variable $TileX$ to tile $P(z_{X})$.  $TileX$ is defined using an affine function $\overrightarrow{\gamma}$ which represents the tiling hyperplanes and the tile size $s$.  $X$ is now an accumulation of the partial answers produced by each tile. 

Additional analysis is needed to tile the reduction body.  Let $R$ be the reduction body of $X$.  We want to tile all the $d$ dimensions of the reduction body using the same approach as mentioned above.  In order to do so, the schedule of tiles must admit the schedules of both variables $R$ and $X$.  Tiling hyperplanes such that the tiling legality constraint (\ref{eq:tilinglegality}) is satisfied can be found.  Reduction body can be partitioned using these tiling hyperplanes to decompose the reduction.

\section{Further improvements}
We identify the following potential areas of improvements to the scheduling techniques presented in this paper.
\begin{enumerate}
\item 
Reiterating over the scheduling approach discussed in section~\ref{sec:solution}, the causality constraints are formulated by making some assumptions regarding $\lambda_{R}$ and then these constraints are simultaneously resolved to find schedules for all the variables.  Hence, while formulating the causality constraints it is not possible to recognize the exact equitemporal hyperplanes without the knowledge of $\lambda_{R}$.  The scheduling techniques do not show how to make an optimal choice of \emph{equitemporal hyperplanes} while formulating the causality constraints.  The analysis is, therefore, sub-optimal and can be improved.

\item 
The suggested scheduling and tiling techniques do not consider the program size parameters.  Reconsider the example in Figure~\ref{fig:tile-eg} (a) with modified size parameters such that $0 \leq i \leq M$ and $0 \leq i \leq N$.  Assume $\lambda_{R}(z_{R}) = i$ as the equitemporal hyperplanes as shown in Figure~\ref{fig:tile-eg} (c).  The number of equitemporal hyperplanes will be given by $M$ and the size of an equitemporal hyperplane will be given by $N$.  The causality constraints in the equation (\ref{eq:guptaslackX}) will impose the condition that the slack must be greater than or equal to the size.  Either $M \geq N$ or $M < N$.  Without the knowledge of the values of $N$ and $M$, it is not possible to find a schedule that satisfies both the inequalities.  Assuming that the values of $N$ and $M$ are known and that $M > N$, after solving if we get $\lambda_{R}(z_{R}) = t = j$ which used $N$ as the size of equitemporal hyperplanes, we get an incorrect schedule of $X$.  Therefore, it becomes necessary to consider size parameters. 

\item 
In the situations where equitemporal hyperplanes have different slacks, it is suggested that equitemporal hyperplanes be scheduled in the decreasing order of slack.  However, the method does not discuss ordering of equitemporal hyperplanes when they have the same slack.

\item 
Furthermore, if the reduction operator is not commutative then accumulations must admit an order.  The scheduling techniques presented in this paper can be extended to consider non-commutative operators.  For example, a reduction computation with an associative but non-commutative operator can be tiled using the techniques presented in Section 5 with additional constraint such that accumulations within a tile are lexicographically ordered.  The accumulations of partial answers into the final answer can also be ordered in the lexicographically increasing order of tiles.  Such an ordering imposed by non-commutative operator might slow down the schedule by a constant factor.  Parallelism can, however, be exploited irrespective of the commutativity of the operator.

\end{enumerate}

\section{Related work}

The scheduling technique of ~\cite{karp1967organization} solved the problem of scheduling Systems of Uniform Recurrence Equations (SUREs).  Using polyhedral ~\cite{Rajopadhye-synth} presented a technique for synthesizing systolic architectures from recurrence rquations which enable scheduling Affine Recurrence Equations. ~\cite{feautrier92a,feautrier92b} give closed form schedules as affine functions of the indices of a nested loop program.  The reader is referred to the book~\cite{DRV-sched00} which details scheduling algorithms for recurrence equations.

The problem of finding schedules in the presence of reductions was initially tackled by~\cite{Redon}.  They assumed a CRCW PRAM model where accumulations can be carried out in a single time step.  They also show that their technique can be extended to machines with a bounded number of processors by serializing reductions using a partial-binary-tree algorithm.  However, they did not show how this can be done efficiently.  Building over their scheduling technique, ~\cite{Gupta} developed an algorithm to determine effective serialization of reductions to achieve the fastest possible linear schedules on an exclusive writes machines with bounded fan-in.  Scheduling SAREs do not need to consider memory based dependence like ~\cite{Polly,Sato2011,uday-pldi08} and hence provide more flexibility.  

While scheduling SAREs with reductions, the reduction dependences are implicit which also allows for maximal parallelism.  Other techniques such as~\cite{pugh-toplas94,Stock2014} make the dependences explicit to improve parallelization.  Automatic Parallelization technique such as Polly's polyhedral optimizer~\cite{Polly} tries to achieve parallelism by introducing privatization.

The problem of finding optimal schedules are directed towards optimizing some cost function that miniminze latency or delay, or maximize fine-grained parallelism~\cite{feautrier92a,feautrier92b,Redon}.  Tiling improves  data locality and works such as ~\cite{uday-pldi08} process loops where reductions are serialized.

In certain cases, it becomes necessary to find piecewice linear schedules. It is, however, difficult to determine the pieces automatically. In the paper~\cite{Mostly-tileable}, the authors show how to find the optimal piece-wise schedule for Optimal String Parenthesization problem and use the \textit{Mostly-Tileable} technique for tiling.  The schedule was, however, found by hand.  
\section{Conclusion}\label{sec:concl}
We studied previous works that address the problem of scheduling SAREs in the presence of reductions.  We show that method of scheduling reductions developed in~\cite{Gupta} has an error in the formulation of causality constraints which leads to concurrent writes.  We exposed this error with an example and provided a solution that guarantees exclusive writes.  The scheduling techniques presented in this paper gives optimal linear schedules.

    Above all, reductions remain memory-bound computations.  Therefore, exploiting data locality using tiling techniques can improve the performance~\cite{WL-pldi91}.  Using the knowledge of reduction operator being associative and commutative, we extended Gupta's scheduling technique to scheduling as well as tiling reductions.  Tiling is also useful for coarse-grained parallelism~\cite{min-sync-lim-lam,Xue00tiling}.  We demonstrated that tiling the equitemporal hyperplanes renders maximal parallelism.  When the accumulations are serialized, like most other techniques do, then similar parallelism can not be achieved because serializing imposes an execution order on tiles.  

The tile size optimization technique presented in this paper maximizes parallelism; employs data locality and provides work efficient solutions which also reduces the total number of synchronizations at the same time. This is achieved by reducing the number of elements that contribute to final accumulation.  



\bibliographystyle{ACM-Reference-Format-Journals}
\bibliography{references}


\end{document}